\documentclass[aps, prb,twocolumn,superscriptaddress, nofootinbib, showpacs]{revtex4-1}

\usepackage{graphicx}
\usepackage{dcolumn}
\usepackage{dcolumn}
\usepackage{textgreek}
\usepackage{textcomp}
\usepackage{adjustbox}

\newcommand{\leqrange}[3]{$#1 \leq\ #2 \leq\ #3$}

\def\KtCA{K\textsubscript{2}Cr\textsubscript{3}As\textsubscript{3}}

\def\ACA{\textit{A}\textsubscript{2}Cr\textsubscript{3}As\textsubscript{3}}
\def\CCA{Cs\textsubscript{2}Cr\textsubscript{3}As\textsubscript{3}}
\def\RCA{Rb\textsubscript{2}Cr\textsubscript{3}As\textsubscript{3}}

\def\a{\textit{a}}

\def\c{\textit{c}}
\def\V{\textit{V}}

\def\Tc{$T_c$}

\def\Na+{Na\textsuperscript{+}}
\def\Ba+{Ba\textsuperscript{2+}}

\def\degrees{$^{\circ}$}
\def\iA{\text{\AA}\textsuperscript{-1}}

\begin{document}

\preprint{APS/123-QED}

\title{Coupling of structure to magnetic and superconducting orders in quasi-one-dimensional  K\textsubscript{2}Cr\textsubscript{3}As\textsubscript{3}}

\author{K.M. Taddei}
\email[corresponding author ]{taddeikm@ornl.gov}
\affiliation{Quantum Condensed Matter Division, Oak Ridge National Laboratory, Oak Ridge, TN 37831}
\author{Q. Zheng}
\affiliation{Materials Science and Technology Division, Oak Ridge National Laboratory, Oak Ridge, TN 37831}
\author{A.S. Sefat}
\affiliation{Materials Science and Technology Division, Oak Ridge National Laboratory, Oak Ridge, TN 37831}
\author{C. de la Cruz}
\affiliation{Quantum Condensed Matter Division, Oak Ridge National Laboratory, Oak Ridge, TN 37831}

\date{\today}

\begin{abstract}
Quasi-one-dimensional \ACA\ (with \textit{A} = K, Cs, Rb) is an intriguing new family of superconductors which exhibit many similar features to the cuprate and iron-based unconventional superconductor families. Yet in contrast to these systems, no charge or magnetic ordering has been observed which could provide the electronic correlations presumed necessary for an unconventional superconducting pairing mechanism - an absence which defies predictions of first principles models. We report the results of neutron scattering experiments on polycrystalline \KtCA\ $(T_c \sim 7\text{K})$ which probed the low temperature dynamics near \Tc . Neutron diffraction data evidence a strong response of the nuclear lattice to the onset of superconductivity while inelastic scattering reveals a highly dispersive column of intensity at the commensurate wavevector $q = (00\frac{1}{2})$ which loses intensity beneath \Tc\ - indicative of short-range magnetic fluctuations.  Using linear spin-wave theory we model the observed scattering and suggest a possible structure to the short-range magnetic order. These observations suggest that \KtCA\ is in close proximity to a magnetic instability and that the incipient magnetic order both couples strongly to the lattice and competes with superconductivity - in direct analogy with the iron-based superconductors.     
 
\end{abstract}

\pacs{74.25.Dw, 74.62.Dh, 74.70.Xa, 61.05.fm}

\maketitle


Understanding how superconductivity arises from myriad competing ground states and exotic phenomena such as quantum criticality has been an overarching theme in the study of unconventional superconductors (UNSC) \textemdash\ particularly in the well-studied quasi-two-dimensional (Q2D) cuprate and iron-based (FBS) families \cite{Basov2011}. Recently, a new family of superconducting quasi-one-dimensional (Q1D) \ACA\ (233) materials (with \textit{A} = K, Cs, Rb) have proven fertile grounds for applying this general narrative to another system which has further lowered-dimensionality \cite{Boa2015,Jiang2015}.  

\begin{figure}
	\includegraphics[width=\columnwidth]{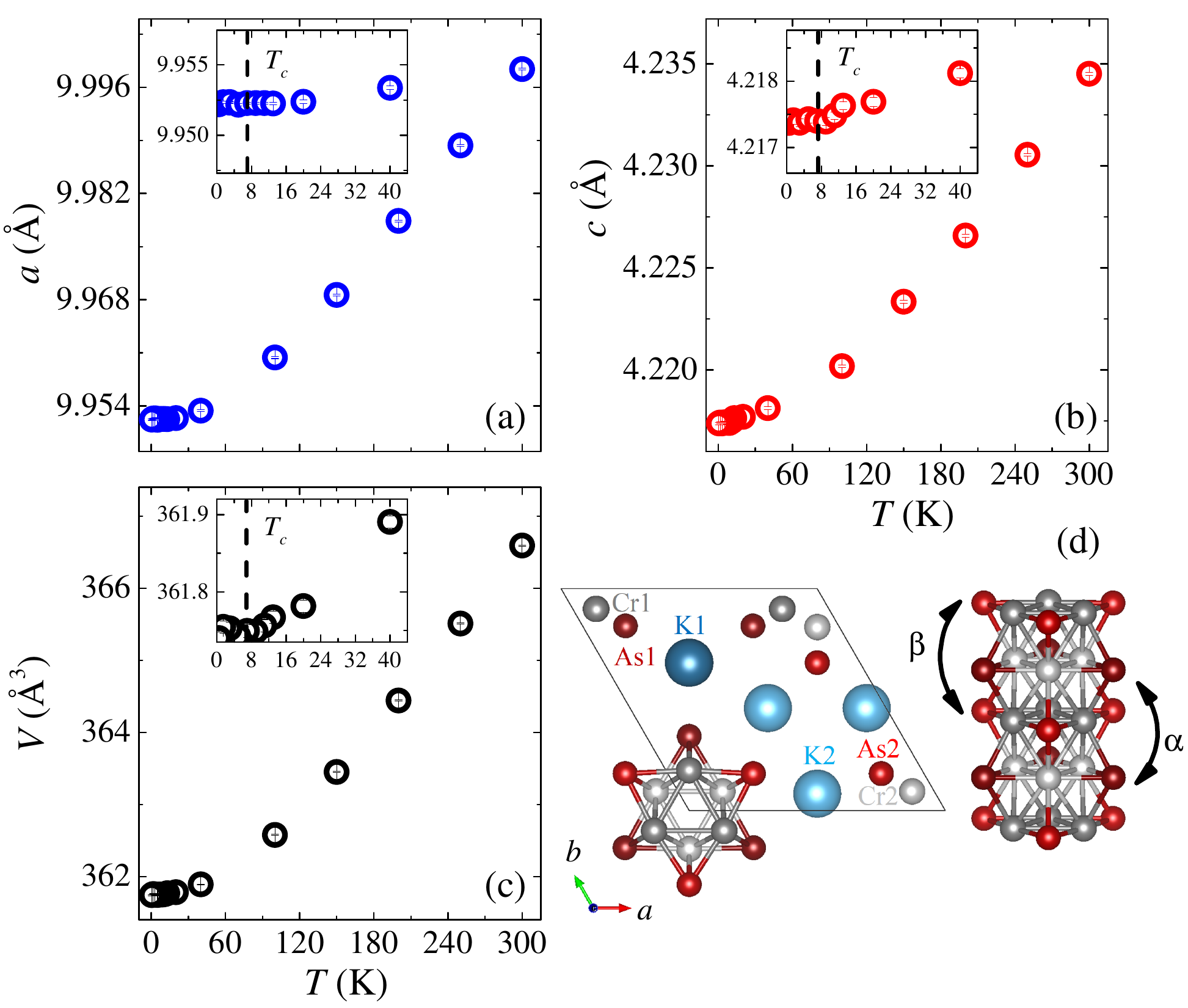}
	\caption{\label{fig:one}  Temperature dependence of \KtCA\ lattice parameters, \textit{a}(a), \textit{c}(b) and the unit cell volume \textit{V}(c), extracted from Rietveld refinements of the NPD patterns. Insets show an expanded view of the low temperature behavior. All \textit{y}-axis ranges show the same relative percent change of the parameters. The crystal structure of \KtCA\ seen along the \textit{c}-axis and the structure of the DWS with angles As2-Cr2-As2 and As1-Cr1-As1 denoted by $\alpha $ and $\beta $ respectively. }	
\end{figure}

The 233 family orders in non-centrosymmetric hexagonal $P\overline{6}m2$ space group symmetry with a structural motif of double-walled sub-nano tubes (DWS) coaxial to the \c -axis and of $[(\text{Cr}_3\text{As}_3)^{-2}]_\infty$ stoichiometry (see Fig.~\ref{fig:one}(d)) while the \textit{A}-site ion acting as a spacer/charge reservoir \lq layer\rq \cite{Boa2015,Tang2015c,Wang2015}. DFT calculations predict a Fermi surface built of complex mixtures of the Cr $3d$ shells with strong Q1D character \cite{Jiang2015,Wu2015,Alemany2015}. Consequently, predictions of Tomonaga-Luttinger liquid (TLL)/general non-Fermi-liquid (nFL) physics, Peierls distortions, ferromagnetic (FM) fluctuations/magnetic ordering and spin-triplet superconductivity have arisen creating a sea of possible ground states and interactions out of which superconductivity stabilizes.  \cite{Jiang2015,Alemany2015,Miao2016,Zhang2016,Zhong2015,Wang2015}. 

Experimentally a similarly complex picture has emerged. nFL behaviors are observed in transport, nuclear magnetic resonance (NMR) and muon spectroscopy ($\mu$SR) measurements, which indicate significant electron correlations and strong magnetic fluctuations\cite{Bao2015,Zhi2015,Adroja2015}. More exotically, measurements of the penetration depth find nodes in the superconducting gap while those of the upper critical field ($H_{c2}$) find $H_{c2}$ to be highly anisotropic, greatly exceed the Pauli-pair-breaking limit, exhibit an in-plane angular dependence and even a possible Fulde-Feerel-Larkin-Ovchinnikov (FFLO) state\cite{Balakirev2015,Zuo2017,Pang2015}. Furthermore, recent angle-resolved photoemission spectroscopy reports find linear behavior of the spectral intensity near the FS indicating possible TLL type physics \cite{Watson2017}. Yet these findings have found divergent explanations ranging from UNSC with spin-triplet or singlet pairing to conventional phonon driven scenarios\cite{Subedi2015,Balakirev2015,Wu2015x}. This ambiguity arises partially from a lack of direct measurements determining the relevant low temperature orders.  Here we investigate the structure and magnetic behavior of \KtCA\ at low temperatures using neutron scattering techniques. 

Synthesis of polycrystalline \KtCA\ was adopted from References~\onlinecite{Bao2015} and \onlinecite{Tang2015c} (as detailed in the supplemental material)\cite{SM}. Neutron powder diffraction (NPD) data were collected using the HB-2A powder diffractometer at the High Flux Isotope Reactor (HFIR) of Oak Ridge National Laboratory (ORNL) using wavelengths ($\lambda $) of 1.54 and 2.41 \AA .  High resolution synchrotron x-ray data were collected at beamline 11BM-B of the Advanced Photon Source (APS) at Argonne National Laboratory (ANL) with $\lambda\ = 0.414$ \AA .  Detailed structural analyses were performed using the Rietveld method as implemented in the FullProf, GSAS and EXPGUI software suites\cite{Rodriguez-Carvajal1993,Toby2006,Larsen2015}. Inelastic neutron scattering (INS) experiments were carried out on HFIR's triple-axis spectrometer HB-3. Tight collimation was used with a fixed analyzer energy of 14.7 meV.    

Our analysis of both the high resolution XRD and NPD patterns found models of \KtCA\ exhibiting the hexagonal space group symmetry $P\overline{6}m2$ to produce the highest quality fits in accord with previous reports (see SM for details)\cite{Bao2015}. Considering our room temperature refinements the crystallographic data agree with those reported by Bao \textit{et. al.}. 

Upon decreasing temperature from 300K to 0.5 K \a , \c\ and \V\ monotonically contract (Fig.~\ref{fig:one}). For temperatures $T > 40$ K a linear thermal expansion is observed (with nearly negligable change for $T<40$ K) with no evidence of a structural response which might arise from nuclear or magnetic orderings in contrast to DFT calculations' predictions but in accord with featureless tansport data \cite{Wu2015,Bao2015,Alemany2015}. While it is possible that the relatively large steps in temperature between 300 K and 20 K might miss a subtle lattice response, such as that commonly reported for the FBS, comparisons of the high temperature and low temperature patterns do not reveal the presence of new peaks, peak splitting or other evidence of a phase transition (see SM) \cite{Avci2013d,Allred2014,Taddei2016}.

Such structural stability, despite a predicted Peierls instability, has also been observed in the Q1D chevrel family (Tl\textsubscript{2}Mo\textsubscript{6}Se\textsubscript{6}) which share the DWS structural motif \cite{Potel1980,Brusetti1989,}. In these materials the DWS sublattice was found to be rigid due to the significant metal-metal bonding and unique geometry of the DWS. By analogue, we suggest the significant inter and intra Cr triangle bonding confounds the expected Peierls distortion which cannot easily lower simultaneously the energies of the inter and intra triangle bonds \cite{Brusetti1989,Alemany2015}. 

Considering the lattice parameters in the range of linear thermal expansion, the coefficient of thermal expansion can be obtained via the expression $\alpha\ = 1/V_0(V_0-V)/(T_0-T)$ (where $V$ can be \a\ or \c ) which finds $\alpha $ for \a , \c\ and \V\ as $1.9\times 10^{-5}$, $1.7\times 10^{-5}$ and $5.5\times 10^{-5}$ K$^{-1}$, respectively. The expansion along the \a\ axis is slightly larger than along \c\ as might be expected from the highly anisotropic quasi-1D nuclear structure. This is in accord with reports on \RCA\ and \CCA\ as well as pressure studies both of which found the \a -axis more sensitive to external or chemical pressure \cite{Wang2015,Tang2015r,Tang2015c,Wang2016}. These values of $\alpha$ are similar to the relatively high values reported for paramagnetic (PM) states of the 11 and various 122 members of the FBS family where strong spin-fluctuations are thought to contribute to the thermal expansion \cite{Rotter2008,Hsu2008,Taddei2015,Bohmer2013,Lortz2003,Miyakawa2003,Miyakawa2005,Miyakawa2001}. As we will show similar spin-fluctuations exist in \KtCA\ and may be responsible for the large $\alpha$ values. 

While no clear signature of a structural phase transition exists, close inspection of the low temperature behavior (\leqrange{0.5\: \text{K}}{T}{20\: \text{K}}) of the \c -axis reveals a response of the lattice near $T_c \sim\ 7$ K (Fig.~\ref{fig:one}(b) inset). In this range, the \c -axis exhibits a contraction just before \Tc\ upon cooling while the \a -axis remains constant.  \V\ undergoes a similar contraction reflecting the suddenly reduced \c -axis (Fig.~\ref{fig:one}(c) inset). The effect is anisotropic being observed only in the ostensibly stiffer \c -axis and appears to correspond to the incipient superconducting transition.

\begin{figure}
	\includegraphics[width=\columnwidth]{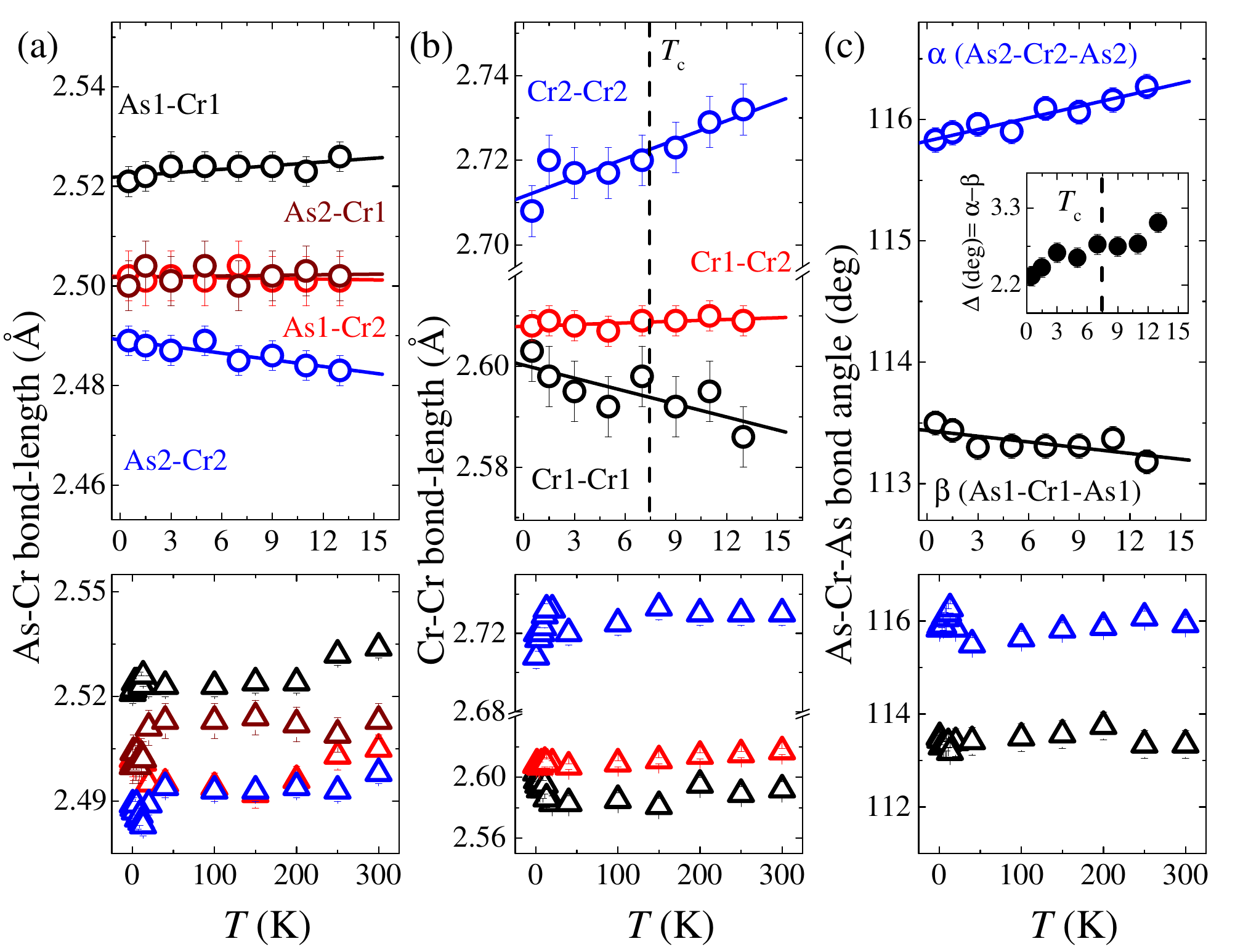}
	\caption{\label{fig:two} Temperature dependence of \KtCA\ As-Cr (a), Cr-Cr (b) and As-Cr-As (c) bond lengths and angle as determined from Rietveld refinements. The upper panels focus on $T < 20\: \text{K}$ while the lower panels show all measured temperatures. Inset of panel (d) shows the degree of non-centrosymmetry as measured by the difference of the two As-Cr-As angles. All scales have been configured to cover similar ranges in percent change of the plotted parameter. Linear fits are provided.}	
\end{figure}

Next we consider the bonding parameters of the DWS and their behavior across \Tc . In general for \leqrange{20\: \text{K}}{T}{300\: \text{K}}, the DWS geometry exhibits little temperature dependence as the As1-Cr1(2) and Cr1-Cr1(2) bond lengths change by $\sim 0.3\%$ while the As2-Cr1(2) and Cr2-Cr2 bond lengths do not change within the sensitivity of our measurements (Fig.~\ref{fig:two}(a) and  (b)). The relatively minor $\sim 0.4\%$ expansion seen in this temperature range along the \c-axis is accounted for by the inter plaquette spacing while the thermal expansion along the \a -axis affects the inter-DWS spacing (via As-K bond lengths). The significant metal-pnictide bond-length rigidity seen here is similar to that reported for FBS (Fig.~\ref{fig:two}(a))\cite{Avci2013d, Taddei2016, Taddei2017}. This is likely due to the strong anti-bonding character of these bonds as determined in Ref.~\onlinecite{Alemany2015}.

Below 20 K an unexpected shift in the DWS bonding behavior is observed. As shown in Fig.~\ref{fig:two}(b), the previously stoic Cr-Cr bond-lengths suddenly exhibit temperature dependence. The Cr1(2)-Cr1(2) bonds dilate(contract) by 0.7\% (0.9\%) as the material is cooled from 13 to 0.5 K - a change larger than twice that from 300 to 20 K. A corresponding contraction (dilation) of 0.2\% (0.24\%) is seen in the As1(2)-Cr1(2) bonds. This describes a sudden adjustment of the intra-CrAs plaquette bonding where the larger Cr2 triangle contracts and the Cr1 triangle dilates while the surrounding As matrix remains rigid (as evidenced by the changing As1(2)-Cr1(2) bond lengths) and the inter-DWS spacing remains unchanged (Fig.~\ref{fig:two}(d)). The previously discussed \c -axis contraction causes a decrease in the inter-plaquette spacing due to the Cr sites' Wyckoff positions' fixed \textit{z} components.  

The cumulative effect is a decrease in the non-centrosymmetry of the DWS tubes, which can be quanitfied (as suggested in Ref~\onlinecite{Wang2016})  as $\Delta = \alpha - \beta$ (where $\alpha$ and $\beta$ denote the As2-Cr2-As2 and As1-Cr1-As1 bond angles respectively Fig.~\ref{fig:one}(d)). Upon decreasuing temperature $\alpha$ closes and $\beta$ widens leading the parameter $\Delta$ to decrease from 3 at 13 K to 2.3 at 0.5 K (Fig~\ref{fig:two}(d)). This is somewhat unexpected in light of pressure effect studies that indicated a positive correlation between \Tc\ and $\Delta$ \cite{Wang2016}. However, we argue that this decrease is a secondary effect due to competition between superconductivity and short-range magnetic order rather than a direct  structural effect of superconductivity.      

\begin{figure*}
	\includegraphics[width=\textwidth]{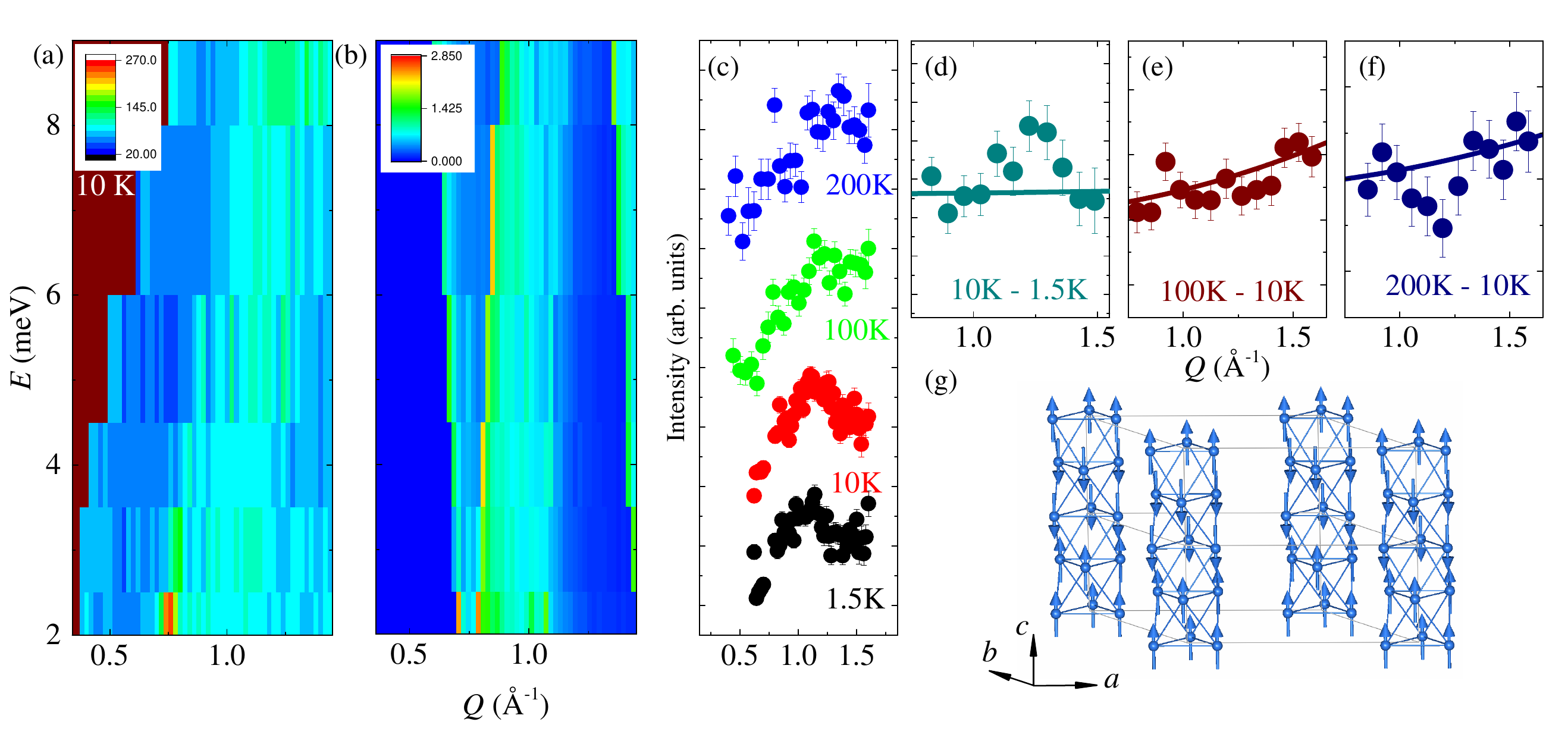}
	\caption{\label{fig:three} Neutron spectra intensity map of the dynamic structure factor with the intensity normalized to the monitor (a). Simulated spin-wave dispersions of the UUDD magnetic structure with arbitrary intensity scale (b). $Q$ dependence of the scattering intensity for $E$ summed over 2-9 meV for all measured temperatures (c) each temperature is offeset by an arbitrary \textit{y} value for visual clarity. Difference curves for the 10 K summed intensity subtracted from the 1.5 K (d), 100 K (e) and 200 K(f) summed intensities. Lines are added to the plots as guides to the eye. UUDD magnetic structure consistent with $Q\sim0.75\: \iA$ (magnetic space group $P_c3c1$) (e).}	
\end{figure*}

To probe for incipient magnetic order INS experiments were performed focusing on low momentum and energy transfers. Fig.~\ref{fig:three}(a) shows a neutron spectra intensity map of the dynamic structure factor $S(Q,E)$ collected at 10 K. Strikingly, a broad column of scattering from \leqrange{0.75\: \iA}{Q}{1.20\: \iA} is clearly seen. The low \textit{Q} and \textit{E} values of this feature indicate that it is not likely due to phonons which exhibit a $Q^{2}$ in $S(Q,E)$ and have a calculated cutoff in \KtCA\ of $\sim 4$ meV \cite{Subedi2015,Osborn2001}. $E$ integrated plots (Fig.~\ref{fig:three}(c)) for data collected at 1.5, 10, 100 and 200 K show a \textit{T}-dependent peak-like feature at $\sim 1.2\: \iA$ visible for $T < 200\: K$. Fig.~\ref{fig:three} (d-f) show subtractions of the \textit{E}-summed data from 10 K. For 100 K $-$ 10 K (Fig.~\ref{fig:three}(e)), no appreciable change in intensity is observed. On the other hand, 200 K $-$ 10 K (panel (d)) exhibits a dip in the intensity at $Q \sim 1.2\: \iA$ indicating that the column of scattering is either not present or has a reduced intensity by 200 K. Similarly, 10 K $-$ 1.5 K (panel (b)) also reveals an intensity difference centered near 1.2 $\iA$. The inelastic signal is suppressed not only at high temperatures but also below \Tc\ - significantly this is consistent with both the results of local probes of the dynamic magnetism in \KtCA\ (nuclear quadrupole resonance (NQR) and Knight shift (KS)) which have suggested short-range magnetic order and with inelastic signals seen of incipient magnetic order in the FBS FeSe and LiFeAs \cite{Zhi2015,Zhi2016,Rahn2015,Taylor2011}.

The $Q$ onset of the column $\sim 0.75\: \iA$ (see Fig.~\ref{fig:three}(a)) is commensurate with a \textit{q} vector $q_m = (00\frac{1}{2})$ which indexes the feature with $hkl$ of $0 0 \frac{1}{2}$ in the nuclear structure. Using representational analysis (as implemented in the ISODISTORT software Ref.~\onlinecite{Campbell2006}) magnetic structures consistent with $q_m=(00\frac{1}{2})$ were explored one of which is shown in Fig.~\ref{fig:three}(g). This model is similar to the \lq up-up-down-down\rq\ (UUDD) magnetic structure predicted in the DFT work of Wu \textit{et. al.} \cite{Wu2015}. The UUDD has Cr moments co-linear with the crystallographic \c -axis and exhibits FM correlations within each plaquette, with FM coupling to the neighboring plaquette in one direction along the chain and AFM along the other (Fig.~\ref{fig:three}(g)).

To test the various magnetic structures consistent with $q_m$, Monte-Carlo simulations were performed using linear spin-wave theory (as implemented in SpinW Ref.~\onlinecite{Toth2015}) to model the observed inelastic scattering as spin-wave dispersions using the exchange interactions (\textit{J}) predicted in Ref.~\onlinecite{Cao2015}. The simulated inelastic powder spectrum of the UUDD magnetic structure reproduces the general features of our experimental spectrum (Fig.~\ref{fig:three}(b))(for more details see SM) though the relatively small region of $S(Q,E)$ covered in our experiment does not allow for a unique determination of the incipient order. In this model, the scattering originates of two acoustic spin-wave modes arising from reflections newly allowed by the UUDD structure: $hkl$ of $00\frac{1}{2}$ and $10\frac{1}{2}$. We note that the general scale of the dispersions appear consistent with the predicted \textit{J} values which put inter-chain exchanges on the order of $\sim 10$ meV and interchain $J$ at $>1$ meV. As the latter of these is increased the branches originating from $hkl$ $00\frac{1}{2}$ and $10\frac{1}{2}$ become distinct and inconsistent with the measurement putting a upper (though non-zero) limit on the strength of the inter-chain coupling. 


This magnetic model together with the observed suppression of the spin-fluctuations below \Tc\ provide a natural explanation for the change in the DWS at \Tc . The sudden changes to the Cr-Cr bonding below \Tc\ are consistent with a reduction of magnetic fluctuations on the Cr sites in the magnetic model shown in Fig.~\ref{fig:three}(d).  In this case, the reduction in the strength of the short-range FM correlations within each plaquette relaxes the Cr-Cr bonds, as was observed in the diffraction data (Fig.~\ref{fig:two}(b)). In Ref~\onlinecite{Wu2015} the different K coordination of the two Cr sites lead the predicted magnetic moment of the Cr2 site to be larger than Cr1. This is consistent with our observation of a stronger response of the Cr2-Cr2 bond length to the reduction of magnetic fluctuations below \Tc\ than the Cr1-Cr1. Furthermore, the sudden reduction of the \c -axis near \Tc\ is consistent with weaking magnetic correlations which reduce the strong AFM next nearest plaquette couplings. This observation elucidates the previous reports of KS and NQR measurements indicating strong short-range magnetic fluctuations below 100 K and agrees remarkably well with recent Raman scattering experiments, which report the sudden softening and hardening of Cr modes at $T < 100\: \text{K}$ prospectively due to strong magneto-elastic coupling \cite{Zhang2015}. Together these results indicate the importance of an incipient magnetic order in the dynamics of this material.

In conclusion, we report results of temperature dependent neutron powder diffraction and spectroscopy experiments. Diffraction data collected between 300 and 0.5 K reveal no signature of a structural phase transition or of the predicted long-range magnetic orders. However, careful inspection of the low temperature thermal expansion shows a significant response of the \c -axis to the onset of superconductivity. Analysis of the internal bonding of the DWS reveal a similar response to \Tc\ where the CrAs plaquettes become less non-centrosymmetric due to a strong contraction of the Cr2-Cr2 bond length. Neutron spectroscopy experiments show the presence of a column of scattering centered at a wave-vector $q_m = 00\frac{1}{2}$ which is suppressed below \Tc\ and exhibits a \textit{Q} dependence consistent with a magnetic origin. We propose that this inelastic signal indicates that \KtCA\ is near a magnetic instability with a tendency to order possibly in an UUDD state. Spin-wave simulations of this structure replicate our observed inelastic signal and generally agree with predictions for the Cr-Cr magnetic interactions. Comparison of the UUDD magnetic structure with the observed bonding behavior at \Tc\ is consistent with competition between short-range magnetic order and superconductivity hinting at a situation similar to that of the FBS. Furthermore, the presence of incipient magnetic order with an AFM $q_m$ lends support to spin-singlet models of superconductivity in these materials and suggests local fluctuations available for mediating electron pairing are AFM in nature.

\begin{acknowledgments}
The part of the research that was conducted at ORNL’s High Flux Isotope Reactor and Spallation Neutron Source was sponsored by the Scientific User Facilities Division, Office of Basic Energy Sciences, US Department of Energy. The research is partly supported by the U.S. Department of Energy (DOE), Office of Science, Basic Energy Sciences (BES), Materials Science and Engineering Division. The authors thank S. Chi for providing help during experimental collection and analysis.
\end{acknowledgments}

\end{document}


\preprint{APS/123-QED}

\title{Supplemental Material for: \\ Coupling of structure to magnetic and superconducting orders in quasi-one-dimensional  K\textsubscript{2}Cr\textsubscript{3}As\textsubscript{3}}

\author{K.M. Taddei}
\email[corresponding author ]{taddeikm@ornl.gov}
\affiliation{Quantum Condensed Matter Division, Oak Ridge National Laboratory, Oak Ridge, TN 37831}
\author{Q. Zheng}
\affiliation{Materials Science and Technology Division, Oak Ridge National Laboratory, Oak Ridge, TN 37831}
\author{A.S. Sefat}
\affiliation{Materials Science and Technology Division, Oak Ridge National Laboratory, Oak Ridge, TN 37831}
\author{C. de la Cruz}
\affiliation{Quantum Condensed Matter Division, Oak Ridge National Laboratory, Oak Ridge, TN 37831}

\date{\today}

\maketitle

\section{\label{sec:synth}Synthesis of K\textsubscript{2}C\MakeLowercase{r}\textsubscript{3}A\MakeLowercase{s}\textsubscript{3}}

\KtCA\ was synthesized following the methodolgy laid out in References~\onlinecite{Bao2015} and \onlinecite{Tang2015c}. High purity elemental K, Cr and As were mixed in the stoichiometric ratio 2:3:3 with an additional 10\% of K to compensate for possible loss to volatilization of the alkali metal. The mixture was placed in an alumina crucible and sealed in an evacuated quartz tube and then slowly heated to 423 K and soaked for 16-20 h. The precursor was made in 2 g batch sizes. After the first annealing cycle the resultant inhomogeneous dark grey powders from the various batches were intimately mixed and pressed into $\sim$ 1 g pellets. 

The final $\sim$ 12 g \KtCA\ product was made in separate 2 g batches. Each batch was made of two pressed pellets of the precursor placed in an alumina crucible and sealed with an arc welder in a steel tube under an atmosphere of Ar. The steel tube was then sealed in an evacuated quartz tube. Each batch was sintered in a box furnace at 973 K for 24 h after a slow ramp. The resultant homogeneous dark grey powder from each 2 g batch was characterized via x-ray diffraction and magnetization measurements to ensure quality before being mixed together into the final 12 g sample. Due to the severe air sensitivity observed in these materials all handling of both precursors and final product was performed in He glove boxes with O and H\textsubscript{2}O concentrations of $ < 0.1$  part per million.   

\newpage

\section{\label{sec:char}Sample Characterization}

\begin{figure}
	\includegraphics{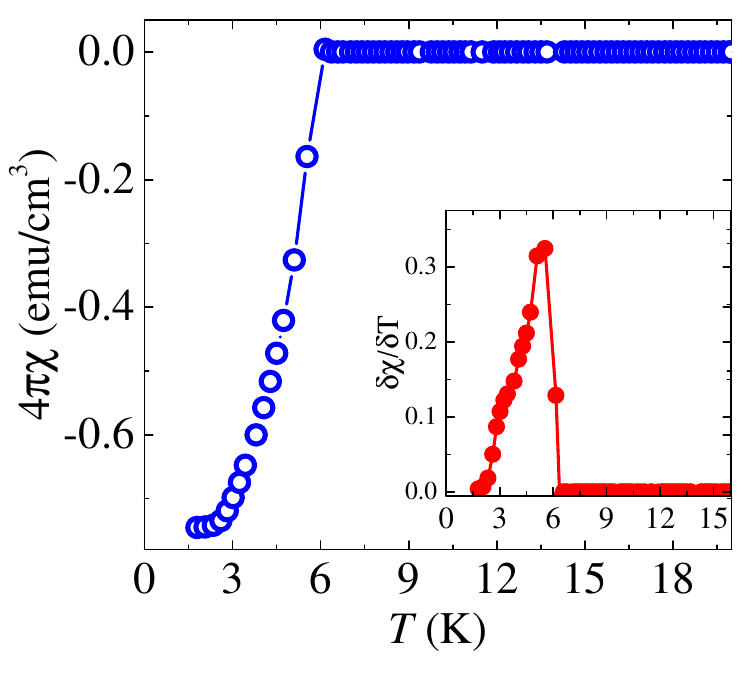}	
	\caption{\label{fig:one}Susceptibility of superconducting \KtCA\ demonstrating a sharp Meissner response (inset shows the $\delta \chi / \delta T$.}	
\end{figure}

Preliminary sample characterization was carried out using a Quantum Design Magnetic Properties Measurement System (MPMS) magnetometer and a Panalytical X'pertPro x-ray diffractometer to determine sample quality. Results of susceptibility measurements are shown in Figure~\ref{fig:one} demonstrating Meissner effect at a sharp superconducting transition \Tc = 6.1 K indicating the high quality and relative chemical homogeneity of our large powder sample.

Analysis of the neutron and x-ray powder diffraction (NPD and XPD) data was performed with Rietveld method as implemented in the FullProf, GSAS and EXPGUI software suites.\cite{Rodriguez-Carvajal1993,Toby2006,Larsen2015} For the continuous wave neutron diffraction data the Thompson-Cox-Hasting formulation for a psuedo-voight peak shape with axial divergnece asymmetry was used \cite{Finger1994}.  A Pseudo-voight peak shape profile function (Type 3 as implemented in GSAS) was used with the synchrotron data. In addition to profile fitting, the atomic positions, thermal parameters of all sites as well as the fractional occupancies of the two K-sties were refined. 

In order to facilitate observation of even the weakest magnetic signal HB-2A's low background low \textit{Q} optimized wavelength ($\lambda = 2.41\: \iA$) was used to collect diffraction patterns between 300 and 10 K. Using a tight collimation and the shorter $\lambda = 1.54\: \iA $ configuration, a second set of diffraction patterns was collected from 17 to 0.5 K. This second set was optimized for tracking subtle crystallographic changes. Therefore, structural parameters for this set of temperatures are reported with a higher level of confidence than the first set. This is reflected in the larger scatter in refined parameters seen in the high temperature range.  

\newpage

\section{\label{sec:Tstruct}Temperature Dependence of Structural Parameters}

\begin{figure}[h]
	\includegraphics[width=\columnwidth]{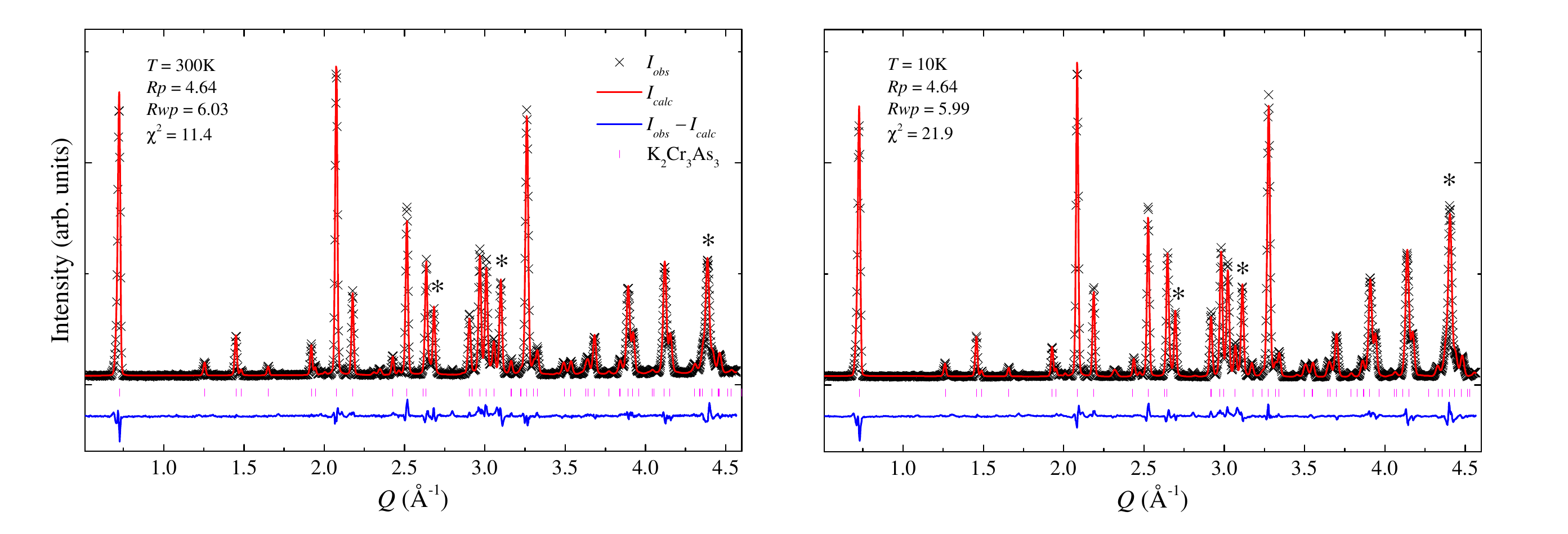}
	\caption{\label{fig:two} Neutron powder diffraction patterns of \KtCA\ collected at 300 and 10 K (left and right panels respectively) with results of Rietveld refinement. Asterisks denote aluminum peaks originating from the sample can.}	
\end{figure}

Figure~\ref{fig:two} shows Rietveld refinements modeling NPD patterns collected at 10 and 300 K on HB-2A with neutrons of incident wavelength $\lambda\ = 2.41$ \AA\ with representative high confidence fits. The space group symmetry $P\overline{6}m2$ produced the highest quality fits as determined by the goodness of fit parameters \textit{Rp} and \textit{Rwp}.  As indicated in Figure~\ref{fig:two} several impurity peaks were observed. While the most significant of these were attributable to the aluminum sample can (indicated by asterisks), small amounts ($>1$\%\ by volume) of minority phases CrAs and Cr\textsubscript{2}As were identified. Special care was taken in all analysis to account for features which might be attributable to these two well studied phases.

\begin{figure}[h]
	\includegraphics[width=\columnwidth]{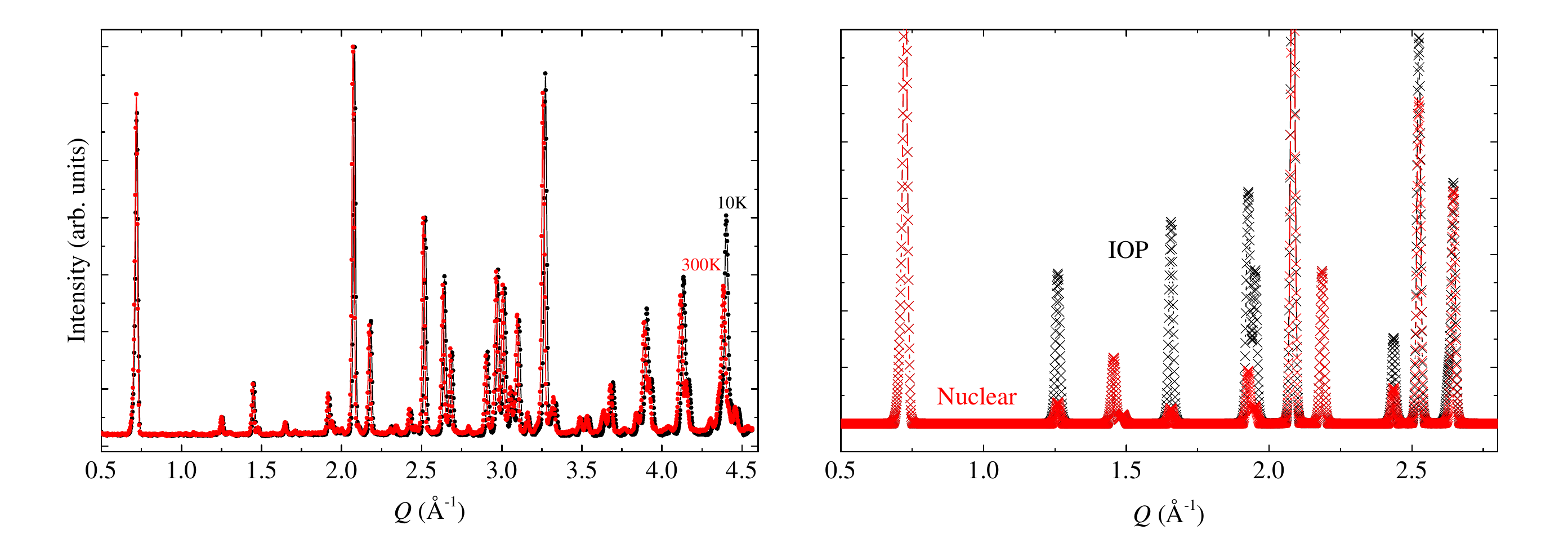}
	\caption{\label{fig:three} Overplot of neutron powder diffraction patterns of \KtCA\ collected at 300 (red) and 10 K (black) (left panel). Simulated powder patterns of the nuclear structure (red) and in-out magnetic structure (IOP) (left panel).}	
\end{figure}

Figure~\ref{fig:three} shows a comparison of the 300 K and 10 K neutron powder diffraction data. As shown, no additional peaks or peak splitting is observed over the entire measured temperature range \leqrange{0.5\: \text{K}}{T}{300\: \text{K}}. Furthermore, modeling of all the magnetic structures predicted from DFT calculations was performed. Fig.~\ref{fig:three} shows a simulated pattern for the in-plane in-out magnetic phase as an example, these phases were found to be inconsistent with the experimentally measured diffraction patterns over all measured temperatures. 

The structural parameters determined from the Rietveld treatments are shown in Table~\ref{tab:one} for NPD patterns collected at 300, 100, 11 and 3 K. Despite significant interest in these materials, few reports of their crystallographic properties have arisen. Most subsequent papers have relied instead on crystallographic data first reported in Ref.~\onlinecite{Bao2015} for \KtCA\ which have become the basis for comparisons with the Cs and Rb compounds as well as for various DFT calculations \cite{Wang2015,Jiang2015,Wu2015}. Our room temperature refinements agree generally with those reported by Bao \textit{et. al.} however, we do see a significant difference in our internal bonding parameters which is persists in refinements using both XPD and NPD data. We attribute this to different K occupancies of our sample (which refined to full K occupancy on both K sites) as compared with Ref.~\onlinecite{Bao2015} (which reported K deficiency).

\begin{table*}[h]
	\caption{\label{tab:one}Structural parameters for \KtCA\ determined at 300, 150, 11 and 3 K.}
	\begin{ruledtabular}
		\begin{tabular}{ldddd}
     		 & \multicolumn{1}{c}{300 K} & \multicolumn{1}{c}{150 K} & \multicolumn{1}{c}{11 K} & \multicolumn{1}{c}{3 K} \\
	\hline
	Space group: $P\overline{6}m2$ & & & & \\
	\a\ (\AA) &  9.9985(2) & 9.9604(1) & 9.95228(8) & 9.95211(8) \\
	\c\ (\AA) &  4.23421(8) & 4.22018(7) & 4.21748(5) & 4.21737(6)\\
	\V\ (\AA$^3$)& 366.57(1) & 362.75(1) & 361.76(1) & 361.75(1)\\
	\hline
	\multicolumn{3}{l}{Bond Lengths (\AA)} & & \\
	As1-As2 & 3.587(2) & 3.578(2) & 3.575(2) & 3.575(2) \\
    As1-K1 & 3.364(3) &	3.354(3) & 3.346(3) & 3.345(3) \\
    As2-K1 & 3.266(3) & 3.251(3) & 3.251(3) & 3.251(3) \\
    As2-K2 & 3.569(3) & 3.545(3) & 3.550(3) & 3.549(3) \\
	As1-Cr1 & 2.529(3) & 2.523(3) & 2.523(3) & 2.524(3) \\
	As1-Cr2 & 2.508(4) & 2.494(4) & 2.501(4) & 2.502(4) \\
	As2-Cr1 & 2.511(5) & 2.513(5) & 2.503(5) & 2.501(5) \\
	As2-Cr2 & 2.500(3) & 2.493(3) & 2.484(3) & 2.487(3) \\
	Cr1-Cr1 & 2.613(5) & 2.585(5) & 2.595(5) & 2.595(5) \\
	Cr1-Cr2 & 2.618(2) & 2.609(2) & 2.610(2) & 2.608(2) \\
	Cr2-Cr2 & 2.718(6) & 2.725(6) & 2.729(6) & 2.717(6) \\
	\hline
	\multicolumn{3}{l}{Bond Angles (\degrees)} & & \\
	$\alpha$ & 115.7(2) & 115.6(2) & 115.5(2) & 116.0(1) \\
	$\beta$ & 113.7(3) & 113.5(3) & 113.5(3) & 113.3(1) \\
	\end{tabular}
	\end{ruledtabular}
\end{table*}

\newpage

\section{\label{sec:GT}Magnetic Structure}

Group theory analysis was performed using the ISOTROPY online software suite \cite{Campbell2006}. Guided by the observed structural stability, magnetic subgroups consistent with the estimated ordering vector ($q_m=(00\frac{1}{2})$, or $A$ $k-\text{point}$) were considered in the absence of any displacement or stain order parameters. The irreducible representations (irrep) associated with ordering vector $q_m$ thus determined are listed in Table~\ref{tab:two}.

\begin{table*}[h]
	\caption{\label{tab:two}Irreducible representations ($\Gamma$), magnetic space groups, magnetic supercell basis vectors and origin for magnetic orderings with $q_m = (00\frac{1}{2})$ }
	\begin{ruledtabular}
		\begin{tabular}{lcccc}
     		 \multicolumn{1}{c}{$\Gamma$} & \multicolumn{1}{c}{Magnetic space group} & \multicolumn{1}{c}{Basis} & \multicolumn{1}{c}{Origin} \\
	\hline
	$mA_1$ & $P_c\overline{6}m2$ &  $(0,-1,0),(1,1,0),(0,0,2)$  & $(0,0,0)$\\
	$mA_2$ & $P_c\overline{6}c2$ &  $(0,-1,0),(1,1,0),(0,0,2)$ & $(0,0,\frac{1}{2})$\\
	$mA_3$ & $P_c\overline{6}m2$ &  $(0,-1,0),(1,1,0),(0,0,2)$ & $(0,0,\frac{1}{2})$  \\
	$mA_4$ & $P_c\overline{6}c2$ &  $(0,-1,0),(1,1,0),(0,0,2)$ & $(0,0,0)$   \\
	$mA_5$ & $A_amm2$ & $(0,0,2),(-1,0,0),(-1,-2,0)$ & $(0,0,0)$   \\
	       & $A_ama2$ & $(0,0,2),(-1,0,0),(-1,-2,0)$ & $(0,0,\frac{1}{2})$\\
		   & $P_bm$   & $(-1,0,0),(0,0,-2),(0,-1,0)$ & $(0,0,0)$ \\
	$mA_6$ & $A_amm2$ & $(0,0,2),(-1,0,0),(-1,-2,0)$ & $(0,0,\frac{1}{2})$   \\
	       & $A_ama2$ & $(0,0,2),(-1,0,0),(-1,-2,0)$ & $(0,0,0)$\\
		   & $P_bm$   & $(-1,0,0),(0,0,-2),(0,-1,0)$ & $(0,0,\frac{1}{2})$ \\
		\end{tabular}
	\end{ruledtabular}
\end{table*}

Starting from the magnetic structures predicted by DFT only the up-up-down-down (UUDD) magnetic structure proposed in Ref~\onlinecite{Wang2015} is consistent with the observed $q_m$.  The UUDD structure is created by a combination of irreps - $ mA_2 \oplus\: mA_4$ - which allows for an active $B_2(a)$ mode on both Cr sites. In the $\overline{6}m2$ point group, only the $B_2(a)$ mode is consistent with magnetic moments co-linear with the \c -axis thus necessitating the combination of two irreps.  

We note that the single irrep magnetic structures with $q_m$ are non-co-linear mixtures of in-plane and out-of-plane modes. Though we are unable to rule-out all such structures due to our small $S(Q,E)$ coverage, we leave it to future studies to determine more rigorously the incipient magnetic structure. In this work we focus on the UUDD structure due to its prediction in DFT and intuitive simplicity.

\section{\label{sec:spin}Spin-wave simulation}

SpinW (Ref~\onlinecite{Toth2015}) was used to model the INS spectra. Our model was built using a two Cr unit cell with the atomic positions and unit cell parameters input as determined from our 10 K structural NPD refinements. The lack of long-range order in our material compounded with our small $S(Q,E)$ coverage contraindicates the use of SpinW to perform meaningful fits. Therefore, we performed modeling using $J$ values determined by the various published DFT treatements of the \ACA\ materials \cite{Jiang2015,Wu2015,Cao2015}. As a further simplification we assume each Cr atom to be in a spin $\frac{1}{2}$ state. In our final calculation of the powder averaged spectrum, 3000 independent orientations were used, $Q$ and $E$ step-sizes were matched to the measured spectra and an energy resolution of 3 meV was adopted. 

We defined exchange interactions for intra-plaquette ($J_1$), inter-plaquette ($J_2$), next nearest plaquette ($J_2'$) and inter-DWS ($J_t$) couplings. The $J$ values predicted by the three DFT studies are: $J_1 = J_2 = 700$ meV; $J_1 = 90$, $J_2 = 60$, $J_2' = -10$ meV; and $J_1 = -9.9$, $J_2 = J_2' = 18.5$, and $J_t = -1$ meV for Ref~\onlinecite{Jiang2015}, \onlinecite{Wu2015} and \onlinecite{Cao2015} respectively. We found the latter of these three to produce the best calculations for the UUDD model and so used them as the starting point. 

Surprisingly, with the UUDD structure the features of the INS spectra discussed in the main text are quite robust. The general features of the calculated powder spectrum are present even when the above parameters are tweaked such as increasing the Cr2 site to spin $\frac{3}{2}$ (as suggested in Ref~\onlinecite{Wu2015}). Overall changes to the scale of the $J$ values affected the level of dispersion as expected, while changes in the ratio of the intra-DWS $J$ were found to quickly destabilize the calculations. However, $J_t$ could be varied continuously without leading to non-physical results. Thus we found that our measured INS spectrum was most consistent with a $J_t$ value lower than that predicted in Ref~\onlinecite{Cao2015} and our final calculation was performed with $J_t = -0.05$ meV.

\begin{figure}[h]
	\includegraphics[scale=.5]{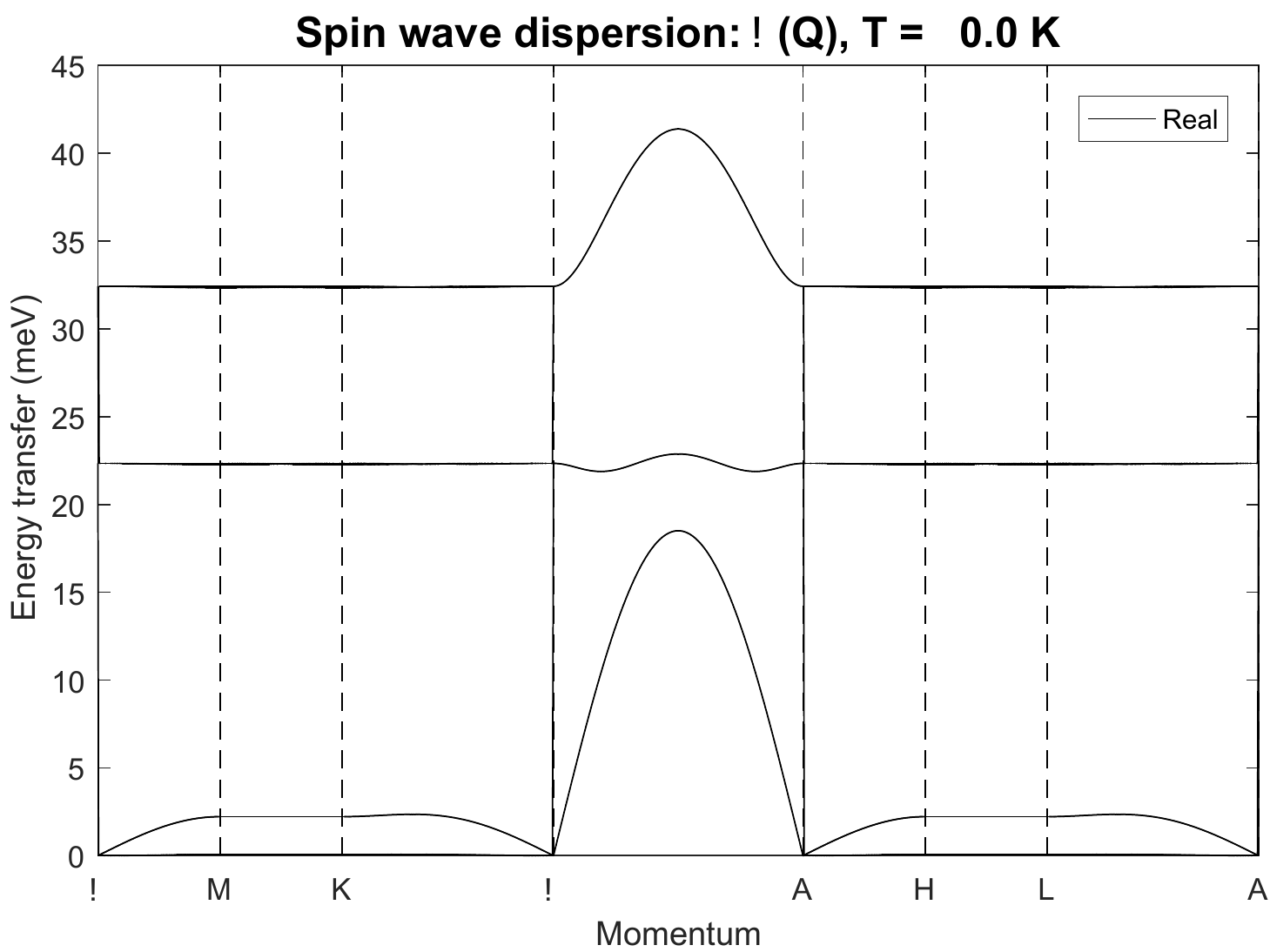}
	\caption{\label{fig:four} Spin-wave dispersions calculated for Cr spin $\frac{1}{2}$, $J_1 = -9.9, J_2 = 18.5, J_2' = 18.5, \text{and} J_t = -0.05$}	
\end{figure}

The resulting non-powder averaged spin-wave dispersions are shown in Fig.~\ref{fig:four}. Predictably the strongest feature is an acoustic branch which shows strong dispersion along the chain direction (going along the reciprocal lattice $c^*$ direction from the Brillioun zone $\Gamma$ point to the $A$ point). Two optical branches are seen at $\sim 23 \text{ and } 33$ meV (each composed of 2 nearly degenerate modes) which show little dispersion for momentum changes in the $a^*b^*$-plane. A weak acoustic mode is seen which shows a non-zero dispersion from in-plane momentum transfer which is due to the non-zero $J_t$ value.

